# Wave-graphene: a full-auxetic carbon semiconductor with high flexibility and optical UV absorption

Linfeng Yu[1], Yi Zhang[1], Jianzhou Lin[1], Kexin Dong[1], Xiong Zheng[1], Zhenzhen Qin[2], and Guangzhao Qin[1,3,4*]

[1] *State Key Laboratory of Advanced Design and Manufacturing Technology for Vehicle, College of Mechanical and Vehicle Engineering, Hunan University, Changsha 410082, P. R. China*
[2] *School of Physics and Microelectronics, Zhengzhou University, Zhengzhou 450001, China*
[3] *Research Institute of Hunan University in Chongqing, Chongqing 401133, China*
[4] *Greater Bay Area Institute for Innovation, Hunan University, Guangzhou 511300, Guangdong Province, China*

**Abstract**

The abundant bonding possibilities of Carbon stimulate the design of numerous carbon allotropes, promising the foundation for exploring structure-functionality relationships. Herein, utilizing the space bending strategy, we successfully engineered a two-dimensional carbon allotrope with pure $sp^2$ hybridization, named "Wave-graphene" from the unique wave-like ripple structure. The novel Wave-graphene exhibits full-auxetic behavior due to anisotropic mechanical response, possessing both negative and zero Poisson's ratios. The fundamental mechanism can be attributed to the fact that highly buckled out-of-plane structures lead to anisotropic responses of in-plane nonlinear interactions, which further lead to anisotropy of lattice vibrations. In addition, Wave-graphene is found having quasi-direct wide bandgap of 2.01 eV, the excellent optical transparency and the high flexibility. The successful design of Wave-graphene with excellent outstanding multifunctional properties shows that the utilization of space bending strategies can provide more degrees of freedom for designing novel materials, further enriching the carbon material family and supplementing its versatility.

---

# Introduction

Carbon, as the foundational element of life, exhibits an extraordinary diversity of allotropes. This diversity arises from carbon's remarkable capacity to form various chemical bonds, encompassing single, double, and triple bonds, leading to a wide array of carbon allotropes. Amid this diversity, graphene [1], a quintessential two-dimensional (2D) carbon allotrope, stands out with its distinctive 2D honeycomb lattice structure comprising six-membered carbon rings. However, the field of carbon allotropes continues to advance, unveiling captivating structures beyond graphene. These structures, such as penta-graphene [2–4], graphdiyne [5–7], graphene+ [8,9], T-graphene [10], biphenylene [11–13], 2D fullerene [14–16] and amorphous carbon monolayer [17], contribute to our understanding of the diverse manifestations of carbon elements. These carbon allotropes offer extensive research domains for scientific exploration and fertile ground for multifunctional structures and materials.

Within this diversity, 2D carbon allotropes showcase a wide range of mechanical properties shaped by their unique structural configurations[18-20]. Among these properties, Poisson's ratio assumes a central role in characterizing their deformation behavior under stress, providing insights into their strain characteristics. As a significant parameter in material mechanics, Poisson's ratio establishes the connection between a material's longitudinal elongation or contraction and transverse contraction or elongation under external forces. It typically ranges from -1 to 0.5 in the classical elastic state [21], although it generally assumes positive values for most materials. However, some materials exhibit a negative Poisson's ratio (NPR). NPR materials offer exceptional properties, including enhanced sound and vibration absorption, increased toughness, resistance to cracking under specific conditions, and elevated shear modulus capabilities [22–25]. These unique properties have sparked great interest in exploring NPR materials in the fields of mechanics such as fasteners, biomedicine, and aerospace [26–28].

Poisson's ratio can display unusual behaviors and unexpected behaviors in 2D materials. When it comes to microscopic mechanisms, a mechanical reentry mechanism is often needed to achieve NPR behavior in a structure [29]. However, the origin of the NPR behavior of 2D materials can be attributed to microscopic electronic behavior caused by low-dimensional quantum effects [30,31]. Furthermore, unique microscopic geometric structures and electronic structures can cause changes in the behavior of NPR, making it show diversity in multiple dimensions, such as bidirectional NPR [32–34], half-NPR [8,35,36], zero Poisson's ratio (0PR) [37–39] and the electrical auxetic effect [40]. However, due to the ubiquitous planar structure and $sp^2$ hybridization of 2D carbon materials, it is challenging to realize NPR with a reentrant mechanism. NPR structures in 2D carbon allotropes are mainly found in materials characterized by $sp^3$ structures, such as penta-graphene [2]. Although NPR can be found in some sp3 hybridized carbons, it still lacks case studies in 2D $sp^2$-hybridized carbon systems.

In this study, utilizing first-principles methodology, we design a novel full-auxetic 2D carbon allotrope, named Wave-graphene (Wave-graphene), as a derivative phase of Janus-graphene. Wave-graphene exhibits lower energy and enhanced stability compared to Janus-graphene. Importantly, in contrast to other sp2-hybridized carbon allotropes, Wave-graphene simultaneously displays both 0PR and NPR behavior, constituting full auxetic behavior. Furthermore, this carbon allotrope possesses wide bandgap semiconductor properties and exceptional optical characteristics. The unique combination of full-auxetic behavior with outstanding electrical and optical properties positions Wave-graphene as a highly promising candidate for multifunctional applications.

**Results and discussion**

**Geometric structure and stability**

Figs. 1(a-h) portray a diverse array of carbon allotropes, encompassing well-known structures like graphene, penta-graphene, grapheneplus, graphdiyne, T-graphene, biphenylene, Janus-graphene, and our novel Wave-graphene. Tables 1 and 2 respectively show the geometric and lattice structure parameters characterizing these eight carbon allotropes, including lattice constants, number of atoms, space group, bonding forms and ring forms. These allotropes can be conceptually understood as rearrangements of carbon rings with 4-, 5-, 6-, 8-, and 12-membered rings in $sp$, $sp^2$ or $sp^3$ hybridized forms. Among them, graphene and five-membered graphene are composed of pure 5-membered rings and 6-membered rings respectively, exemplifying well-established carbon frameworks. In contrast, the remaining allotropes exhibit intricate hybrid compositions of carbon rings, rendering them highly intriguing from a structural perspective. Notably, Wave-graphene, closely related to biphenylene and Janus-graphene, displays a distinctive structural configuration characterized by the presence of 4-6-8 carbon-membered rings. The top view of Wave-graphene closely parallels that of Janus-graphene. Nevertheless, the defining feature of Wave-graphene is its prominent buckling structure in the side view, which notably exhibits a pronounced central symmetry. This unique structural feature designates Wave-graphene as a derivative phase of Janus-graphene, specifically characterized as a non-Janus phase. Moving to the dynamic stability assessment, as illustrated in Fig. 1(i), *ab initio* molecular dynamics (AIMD) simulations[42] were employed to meticulously investigate the time evolution of each atom's energy within Wave-graphene and Janus graphene, conducted at an ambient temperature

of 300 K. During the simulation, the energy of Wave-graphene remained remarkably stable, signifying an absence of atomic bond formation or rupture throughout the dynamic evolution[43], as shown in the embedded part shown in Fig. 1(i). Over the course of 1,000 femtoseconds, the energy per atom in Wave-graphene hovered around -8.67 eV, while the average energy per atom in Janus graphene remained around -8.47 eV. The energy curves of both materials remained stable without discernible fluctuations, further substantiating the structural stability of Wave-graphene and Janus-graphene. Intriguingly, the structural stability of highly buckled Wave-graphene exceeds that of Janus graphene, mainly due to its lower energy, as shown in Fig. 1(j). The dynamic stability of Janus graphene and W graphene was further revealed by studying phonon dispersion, as shown in Fig. 1(k). Their lowest phonon frequency is greater than zero which means their structure is resilient to dynamic perturbations, *i.e.* dynamically stable. To put it briefly, we designed a brand-new class of $sp^2$ hybridized 2D carbon allotrope known as "Wave-graphene" because of its buckling configuration that resembles the letter "W" or its ripple structure that resembles waves. It has been demonstrated theoretically to be stable as a low-energy derivative phase of Janus-graphene.

Table 1. Geometric and structural parameters and related physical properties of graphene, Penta-graphene, graphdiyne and grapheneplus. Lattice constants *a*, *b* and *h*, space group, number of atoms, hybridization form, ring type, elastic constant, Young's modulus, shear modulus, Poisson's ratio and electronic band gap. The "min" and "max" represent the minimum and maximum values respectively.

|  | Graphene | Penta-graphene | Graphdiyne | Grapheneplus |
|---|---|---|---|---|
| *a* (Å) | 2.47 | 3.641 | 6.887 | 6.637 |
| *b* (Å) | 2.47 | 3.641 | 6.887 | 6.637 |
| *h* (Å) | 0 | 1.11 | 0 | 1.24 |
| Space Group | P6/mmm | P-42_1m | P2/m | P4/nmm |
| Number of atoms in the primitive cell | 2 | 6 | 12 | 18 |
| Hybrid form | $sp^2$ | $sp^2$, $sp^3$ | $sp$, $sp^2$ | $sp^2$, $sp^3$ |
| Ring type | 6 | 5 | 6,12 | 5,8 |
| $C_{11}$ (N/m) | 345.275 | 272.713 | 203.635 | 143.758 |
| $C_{12}$ (N/m) | 75.657 | -20.628 | 85.545 | 58.012 |
| $C_{44}$ (N/m) | 134.809 | 38.379 | 59.045 | 30.304 |
| $Y_{min}$ (N/m) | 328.697 | 117.682 | 167.699 | 93.217 |
| $Y_{max}$ (N/m) | 328.697 | 271.153 | 167.699 | 120.348 |
| $G_{min}$ (N/m) | 134.809 | 38.379 | 59.045 | 30.304 |
| $G_{max}$ (N/m) | 134.809 | 146.671 | 59.045 | 42.873 |
| $v_{min}$ | 0.219 | -0.076 | 0.42 | 0.404 |
| $v_{max}$ | 0.219 | 0.533 | 0.42 | 0.538 |
| Bandgap (eV) | - | 3.3 | 0.9 | - |

Table 2. Geometric and structural parameters and related physical properties of T-graphene, biphenylene, Janus-graphene and Wave-graphene. Lattice constants a, b and h, space group, number of atoms, hybridization form, ring type, elastic constant, Young's modulus, shear modulus, Poisson's ratio and electronic band gap. The "min" and "max" represent the minimum and maximum values respectively.

|  | T-graphene | Biphenylene | Janus-graphene | Wave-graphene |
|---|---|---|---|---|
| $a$ (Å) | 3.445 | 4.526 | 5.485 | 7.257 |
| $b$ (Å) | 3.445 | 3.755 | 5.485 | 7.257 |
| $h$ (Å) | 0 | 0 | 1.09 | 3.08 |
| *Space Group* | P4/mmm | Pmmm | P4mm | P4/nmm |
| *Number of atoms in the primitive cell* | 4 | 6 | 12 | 24 |
| *Hybrid form* | $sp^2$ | $sp^2$ | $sp^2$ | $sp^2$ |
| *Ring type* | 4,8 | 4,6,8 | 4,6,8 | 4,6,8 |
| $C_{11}$ (N/m) | 301.478 | 237.818 | 173.788 | 100.646 |
| $C_{12}$ (N/m) | 54.426 | 95.796 | 0.824 | 4.314 |
| $C_{44}$ (N/m) | 49.663 | 82.896 | 81.984 | 87.017 |
| $Y_{min}$ (N/m) | 155.309 | 204.523 | 169.123 | 100.461 |
| $Y_{max}$ (N/m) | 291.652 | 237.037 | 173.784 | 130.946 |
| $G_{min}$ (N/m) | 49.663 | 79.956 | 81.984 | 48.166 |
| $G_{max}$ (N/m) | 123.526 | 82.896 | 86.482 | 87.017 |
| $\nu_{min}$ | 0.181 | 0.347 | 0.005 | -0.248 |
| $\nu_{max}$ | 0.564 | 0.403 | 0.031 | 0.043 |
| Bandgap (eV) | - | - |  | 2.12 |

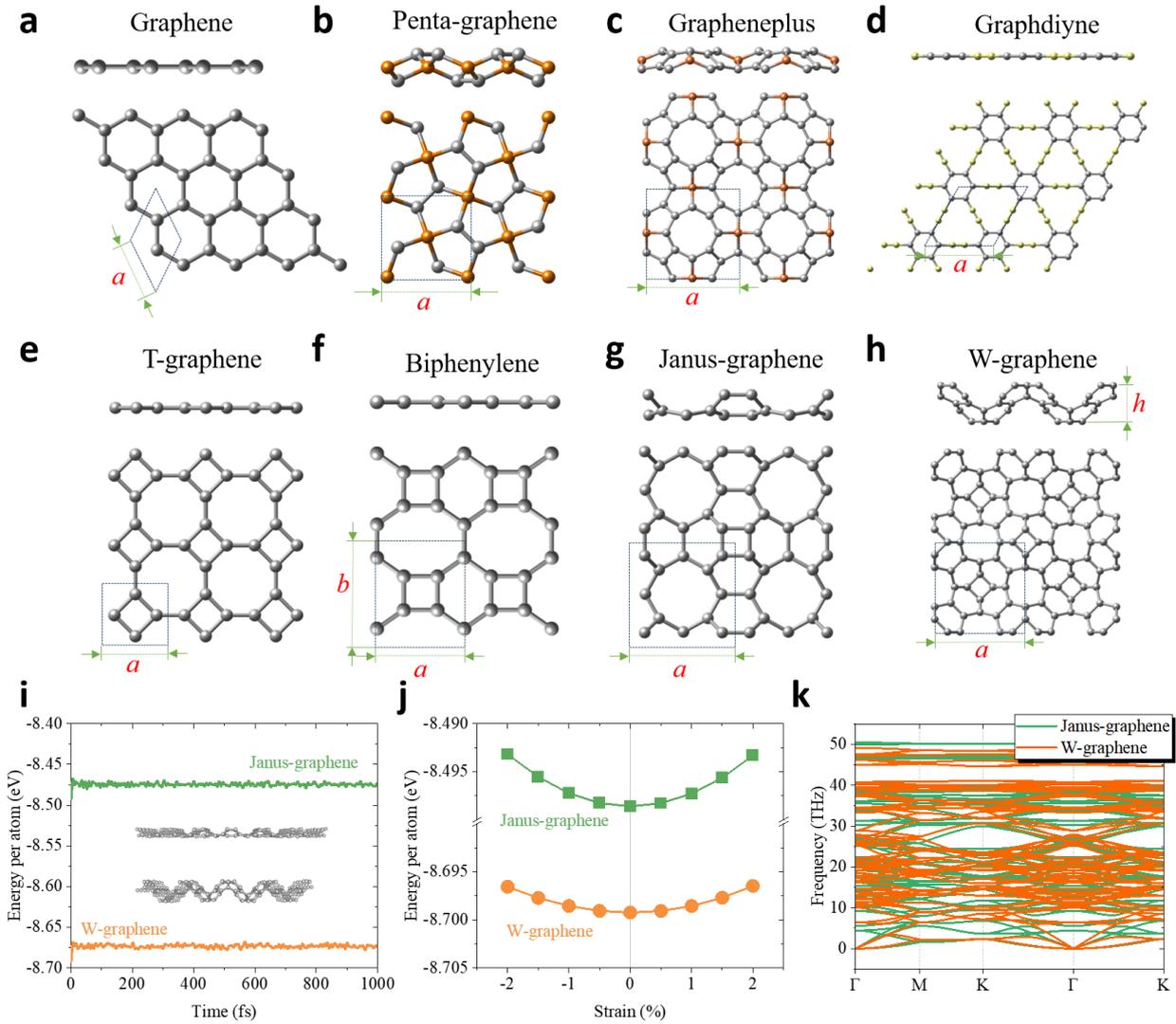

Figure 1. **Geometric structure and stability.** The crystal structures of (a) graphene, (b) Penta-graphene, (c) grapheneplus, (d) graphdiyne, (e) T-graphene, (f) biphenylene, (g) Janus-graphene, and (h) Wave-graphene. Yellow, gray, and brown atoms represent $sp$, $sp^2$, and $sp^3$ hybridized forms, respectively. (i) The dynamic simulation at 300K for Wave-graphene. (j) Comparison of the energy of each atom in the system. (k) Phonon dispersion.

**Electrical properties**

The distinctive atomic arrangement gives rise to a range of diverse physical properties. As depicted in Fig. 2(a), the contributions from various orbitals, such as $s$, $p_x$, $p_y$, and $p_z$, to the energy bands are discerned. The corresponding electronic partial state density is shown in Fig. 2(b) and juxtaposed with Janus-graphene and graphene. In pristine planar graphene, there is pure $sp^2$ hybridization, with $p_z$ orbitals playing a dominant role at the top of the valence band and the bottom of the conduction band. In contrast, due to the obvious out-of-plane buckling in Wave-graphene, it exhibits a dangling bond form similar to $sp^3$ hybridization, even more obvious than in Janus-graphene. Consequently, the valence band top in Wave-graphene is primarily a result of the collective contributions of $p_x$, $p_y$, and $p_z$ orbitals. This unique dangling bond structure is further unveiled through the analysis of the electronic localization function (ELF), as illustrated in Fig. 2(c). The concept of the ELF was initially introduced by *Becke* and *Edgecombe* in 1990 [44]. Beginning with a given wave function, it defines a spatial region that correlates with electron pairs[45], measuring the likelihood of locating electrons near another electron of the same spin, which is expressed as [44]:

$$\text{ELF} = \left(1 + \left\{\frac{K(r)}{K_h[\rho(r)]}\right\}^2\right)^{-1}, \qquad (1)$$

where $K(r)$ is the true electron gas density, and $K_h[\rho(r)]$ is the uniform electron gas density. ELF values range between 0 and 1. An ELF value of 1 indicates complete electron localization in that region, while an ELF value of 0.5 signifies electron distribution akin to an electron gas in that area. An ELF value of 0 suggests complete electron delocalization or an absence of electrons in the region. Owing to the significant out-of-plane buckling in Wave-graphene, there is a redistribution of electron density in the out-of-plane direction, resulting in an enhanced coupling between $p_z$ orbitals and $p_x$ and $p_y$ orbitals[46]. This augmented coupling manifests an electronic band structure akin to $sp^3$ hybridization, even though the atomic configuration maintains a $sp^2$ hybridized structure. Changes in electronic structure are easily discernible in ELF and depict a highly asymmetric ELF distribution in the out-of-plane direction[47], which further destroys the ideal distribution of large cyclic π bonds in the $p_z$ orbitals, a characteristic of pristine graphene. Hence, Wave-graphene exhibits semiconductor properties, while pure $sp^2$ hybridized graphene and biphenylene exhibit semi-metallic properties.

The PBE functional may underestimate the band gap of a material in some cases due to its approximate nature, especially for insulators and semiconductors. In order to obtain the accurate bandgap, the HSE06 functional was executed to evaluate the band structure of Wave-graphene[48], as shown in Fig. 2(d). The conduction band minimum (CBM) of Wave-graphene is at the high symmetry point Γ, while the valence band maximum (VBM) is at the high symmetry point path K-Γ, revealing the semiconductor characteristics of the indirect bandgap. However, the valence band values exhibit flat-band characteristics, leading to a band gap of 2.12 (1.45) eV at Γ point for the HSE06 (PBE) functional. As shown in Fig. 2 (e), under the flat band characteristics, the difficulty of electron transitions from points on VBM1 and other highly symmetric paths (such as VBM2) to CBM is almost similar, leading to the quasi-direct semiconductor properties of Wave-graphene. Fig. 2 (f) compares the electronic band gaps of eight 2D carbon allotropes. The electronic bandgap of Wave-graphene is similar to Janus graphene, but slightly lower than penta-graphene (3.3 eV) and higher than graphene (0 eV), grapheneplus (0 eV), and graphdiyne (0.9 eV). Thus, Wave-graphene exhibits wide-bandgap semiconductor properties with a quasi-direct bandgap.

**Optical properties**

The quasi-direct electronic properties of Wave-graphene with a wider bandgap can further bring excellent optical properties. The optical properties can be derived from frequency-dependent dielectric function with the real $\varepsilon_1(\omega)$ and imaginary $\varepsilon_2(\omega)$ parts [49] :

$$\varepsilon(\omega) = \varepsilon_1(\omega) + i\varepsilon_2(\omega) . \tag{2}$$

The imaginary part $\varepsilon_2(\omega)$ can be further expressed as: [50]

$$\varepsilon_2(\omega) = \frac{4\pi^2 e^2}{m^2 \pi^2} \times \sum_{C,V} |P_{C,V}|^2 \delta(E_C + E_V - \hbar\omega) , \tag{3}$$

where *e*, m, *ω* are the electronic charge, the effective mass, and the angular frequency, respectively. $P_{C,V}$ represents the momentum conversion matrix between the conduction band *C* and valence band *V*. *E* is the electronic energy level. The delta function (*δ*) can derive the energy conversion of electrons during the transition from band to band. $\varepsilon_2(\omega)$ should be normalized with the vacuum layers: [51]

$$\langle \varepsilon_2 \rangle = \frac{L}{d} \varepsilon_2 , \tag{4}$$

where *L* and *d* respectively represent the layer spacing and effective thickness. Simultaneously, $\langle \varepsilon_1 \rangle$ can be expressed based on the Kramers-Kronig dispersion relation: [52]

$$\langle \varepsilon_1 \rangle = 1 + \frac{2}{\pi} \int_0^\infty \frac{\omega' \langle \varepsilon_2(\omega') \rangle}{\omega'^2 - \omega^2} d\omega' . \tag{5}$$

Furthermore, the linear photon spectrums are obtained based on the dielectric function. The absorption coefficient α($\omega$) can be calculated as: [49]

$$\alpha(\omega) = \frac{\sqrt{2}\omega}{c} \left[ \frac{\sqrt{\varepsilon_1(\omega)^2 + \varepsilon_2(\omega)^2} - \varepsilon_1(\omega)}{2} \right]^{\frac{1}{2}}. \tag{6}$$

When photons possess energies higher than the electronic band gap of the semiconductor, wide-bandgap semiconductors readily absorb these photons, resulting in the generation of electron-hole pairs. This property finds broad application in solar cells and photodetectors, as it allows for the efficient conversion of light energy into electrical energy. For Wave-graphene, photons with energy higher than 2.12 eV are easily absorbed. As shown in Fig. 2(g), the optical absorption peak is significantly amplified above 2 eV and reaches the maximum peak at approximately 4 eV. Wide bandgap semiconductors typically demonstrate significant transparency within the visible light spectrum. This transparency results from the electronic band gap (bandgap energy) of wide bandgap semiconductors surpassing the energy of visible light photons. When the photon energy falls below or matches the band gap energy of the semiconductor, it lacks the necessary energy to elevate electrons from the valence band to the conduction band within the semiconductor. Consequently, the photon remains unabsorbed and is either reflected or transmitted by the semiconductor material, imparting transparency to visible light. Visible light generally spans an energy range between 2 eV and 3.1 eV, representing the colors of the rainbow. In contrast, wide-bandgap semiconductors typically possess bandgap energies greater than this range. As illustrated in Figs. 2(h-i), the electronic band gap of Wave-graphene measures 2.12 eV, indicating its transparency to light with wavelengths exceeding 2.12 eV, particularly when the refractive contribution is minimal (not more than 1% in Fig. 2(i)). This property makes Wave-graphene suitable as an optical window material or as a component in transparent electronics within the visible light range. Therefore, the strong photon absorption capabilities of wide bandgap semiconductors for higher-energy photons and their transparency to visible light render them valuable materials for various applications. These applications range from efficient light-to-electricity conversion in solar cells and photodetectors to serving as optical window materials for transparent electronics.

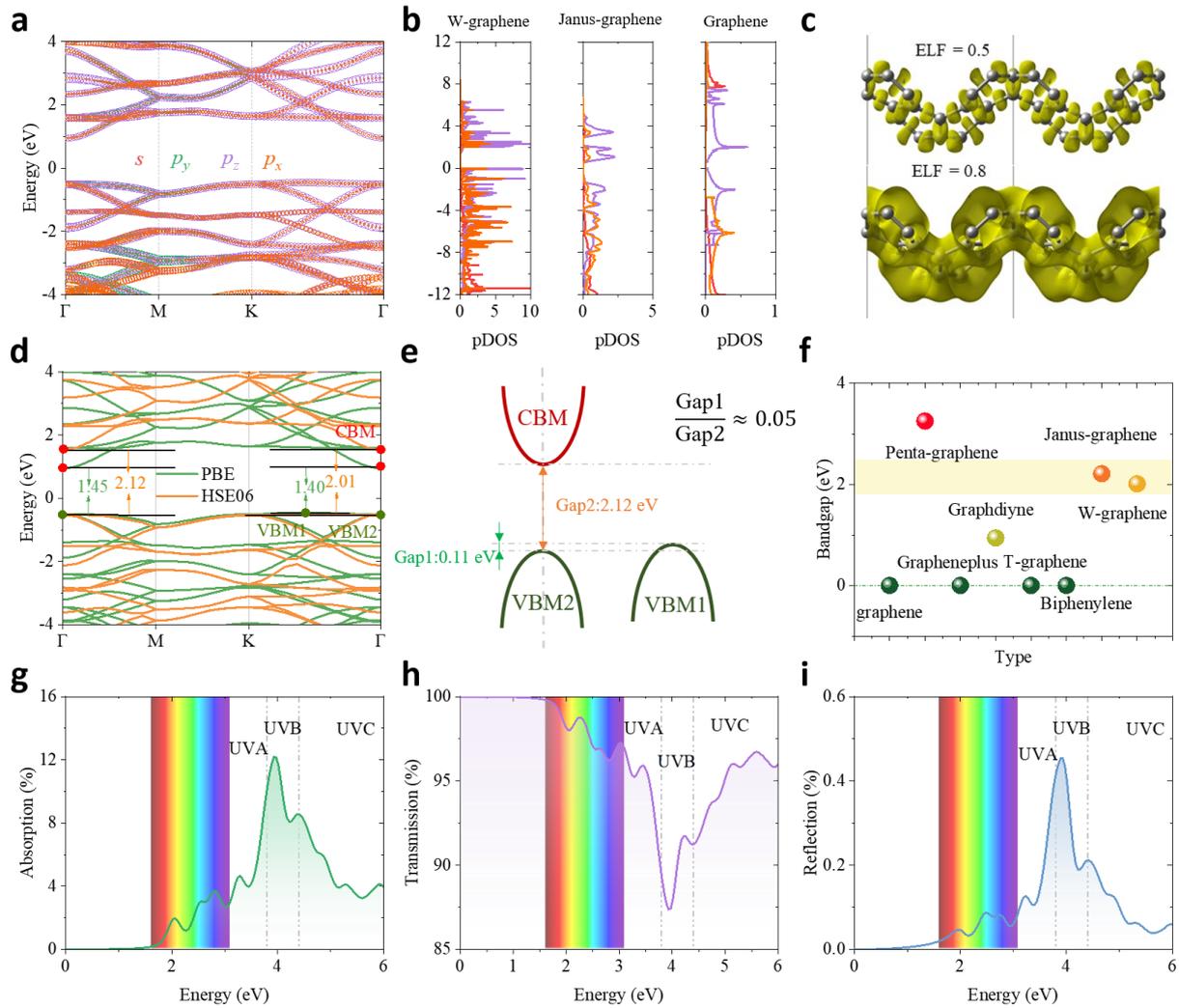

Figure 2. **Electronic properties.** (a) The contribution of different electron orbitals to energy bands. (b) Partial density of states for Wave-graphene, Janus-graphene, and graphene. (c) Electron localization function for Wave-graphene. (d) The electronic bandgap calculated by the HSE06 functional. (e) Schematic diagram of VBM and CBM. (f) Electronic band gaps of different carbon allotropes. (g) The optical absorption, (h) transmission, and (i) reflection coefficients of Wave-graphene.

**Young's modulus and shear modulus**

The mechanical stability of materials can be evaluated based on the elastic matrix derived from Hooke's law [53]:

$$\begin{bmatrix} \sigma_{xx} \\ \sigma_{yy} \\ \sigma_{xy} \end{bmatrix} = \begin{bmatrix} C_{11} & C_{12} & C_{16} \\ C_{12} & C_{22} & C_{26} \\ C_{61} & C_{62} & C_{66} \end{bmatrix} \begin{bmatrix} \varepsilon_{xx} \\ \varepsilon_{yy} \\ 2\varepsilon_{xy} \end{bmatrix}, \tag{7}$$

where the standard Voigt notation of 1-*xx*, 2-*yy*, and 6-*xy* is used to mark elastic constant $C_{ij}$ (*i, j* = 1, 2, 6). Further, The $C_{ij}$ is obtained by the second derivative of strained energy using on energy-strain method [54]:

$$U(\varepsilon) = \frac{1}{2} C_{11}\varepsilon_{xx}^2 + \frac{1}{2} C_{22}\varepsilon_{yy}^2 + C_{12}\varepsilon_{xx}\varepsilon_{yy} + 2C_{66}\varepsilon_{xy}^2. \tag{8}$$

The mechanical stability can be judged by Born-Huang criteria, which requires the elastic constant to comply with the relationship: $C_{11}C_{22} - C_{12}^2 > 0$ and $C_{66} > 0$. Young's modulus $Y(\theta)$, Poisson's ratio $v(\theta)$ shear modulus $G(\theta)$ can be calculated by:

$$Y(\theta) = \frac{C_{11}C_{12} - C_{12}^2}{C_{11}\sin^4\theta + C_{22}\sin^4\theta + A\sin^2\theta\cos^2\theta}, \tag{9}$$

$$v(\theta) = \frac{C_{12}\sin^4\theta - B\sin^2\theta\cos^2\theta + C_{12}\cos^4\theta}{C_{11}\sin^4\theta - A\sin^2\theta\cos^2\theta + C_{22}\cos^4\theta}, \tag{10}$$

$$G(\theta) = \frac{Y(\theta)}{2(1 + v(\theta))}, \tag{11}$$

where A = $(C_{12}C_{22} - C_{12}^2)/C_{66} - 2C_{12}$ and B = $C_{11} + C_{22} - (C_{12}C_{22} - C_{12}^2)/C_{66}$.

Young's modulus is a physical quantity that quantifies the stiffness or elastic properties of a material. It expresses the relationship between stress and strain when the material undergoes stretching or compression. Specifically, Young's modulus gauges the extent of elastic deformation a material experiences after being subjected to a load, signifying the material's capacity to revert to its original shape following stress application. A higher (lower) Young's modulus typically indicates that the material is harder or possesses greater (lesser) stiffness. Materials with a high Young's modulus exhibit smaller strains under identical stress levels, implying that their molecular or atomic structures are relatively resistant (more susceptible) to deformation caused by external stress. Consequently, an increase in Young's modulus corresponds to enhanced material stiffness and elasticity, rendering it more resistant to deformation. Conversely, materials with a lower Young's modulus deform more readily. In parallel, materials characterized by a high shear modulus, which describes a material's resistance to shear stress, demonstrate enhanced capability to withstand shear stress. As a result, they are less prone to shear deformation when subjected to shear forces.

As illustrated in Figs. 3(a) and (b), Wave-graphene exhibits comparable Young's modulus and shear modulus to Penta-graphene and Grapheneplus with $sp^3$ hybridized carbon architecture at higher degrees of structural buckling. However, these values are much lower than those of other carbon allotropes with planar architecture. In comparison to Janus-graphene, which shares the same $sp^2$ hybridization and carbon-membered rings, Wave-graphene demonstrates significantly lower Young's modulus and shear modulus, along with a more pronounced anisotropy. Referring to Figs. 3(c-d) and (e-f), the maximum Young's modulus and shear modulus of Wave-graphene are 130 N/m and 85 N/m, respectively. These values are lower than the corresponding maximum Young's modulus (170 N/m) and shear modulus (86 N/m) of Janus-graphene. Furthermore, the minimum Young's modulus and shear modulus of Wave-graphene (105 N/m and 50 N/m) are also lower than those of Janus-graphene (169 N/m and 82 N/m), highlighting the heightened anisotropy in Wave-graphene compared to Janus-graphene. The anisotropy of Wave-graphene, calculated as the ratio of the characterized maximum value to the minimum value, is 1.24 for the Young's modulus and 1.70 for t shear modulus. These values surpass the anisotropy of Janus-graphene, which stands at 1.01 for Young's modulus and 1.05 for shear modulus. This discrepancy is attributed to the increased degree of buckling in Wave-graphene, deepening the anisotropy of atomic arrangement and resulting in more pronounced anisotropic responses in mechanical properties. In short, Wave-graphene's high buckling structure makes it a more flexible 2D carbon material.

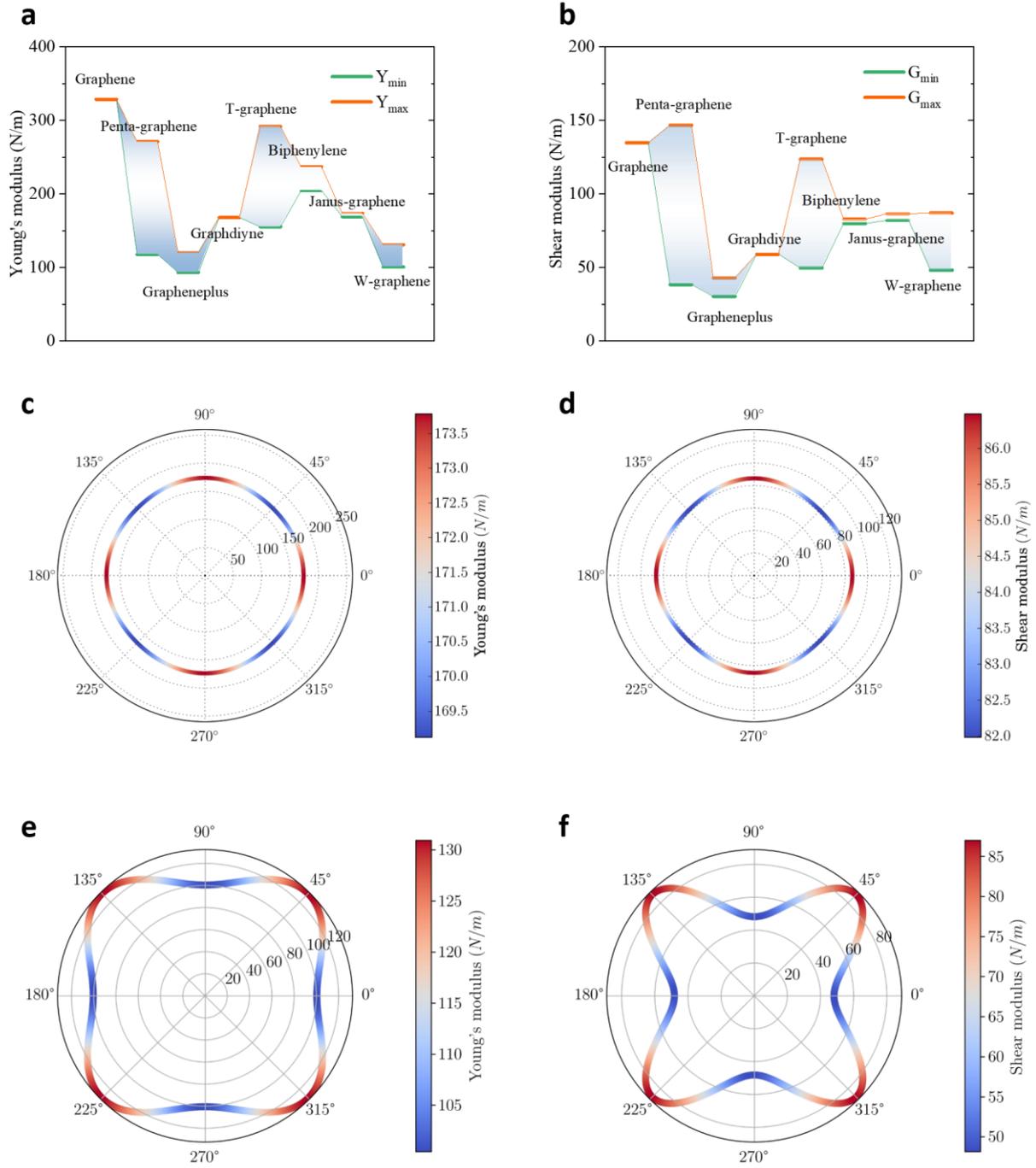

Figure 3. **Mechanical properties.** Comparison of maximum and minimum values of (a) Young's modulus, (b) shear modulus for eight carbon allotropes. The (c) Young's modulus, and (d) shear modulus in different directions for Janus-graphene. The (e) Young's modulus, (f) shear modulus in different directions for Wave-graphene.

**Full-auxetic properties**

Poisson's ratio is a parameter that describes the deformation behavior of a material. When a material is exposed to external tensile stress, Poisson's ratio is frequently positive, as seen in Fig. 4(a), suggesting that the material would compress transversely and extend longitudinally. It is possible for Poisson's ratio to equal zero, indicating that there is essentially no associated deformation in the longitudinal and transverse directions when the material is strained. In some special cases, materials may have a NPR. When a material with a NPR is subjected to external stress, it will not only extend longitudinally along the stress direction, but also expand transversely. Intriguingly, buckling and anisotropic atomic arrangement extend the Poisson's ratio range of Wave-graphene, making it a full-auxetic material spanning negative (NPR), and zero (0PR) Poisson's ratio behavior. As shown in Fig. 4(b), among the eight allotropes, only penta-graphene and Wave-graphene have a negative minimum Poisson's ratio, and Wave-graphene's NPR can reach -0.25, which is much higher than pentagonal graphene. This also indicates that Wave-graphene has strong anisotropy for NPR behavior. Furthermore, the angle-dependent Poisson's ratio behavior of Janus-graphene and Wave-graphene in different directions is shown in Figs. 4(c) and (d). The Poisson's ratio of Janus-graphene ranges between 0.00-0.03, which indicates that the Poisson's ratio of Janus-graphene is almost independent of strain and direction, *i.e.*, 0PR. The degree of buckling of Wave-graphene is higher than that of Janus-graphene, and the anisotropy of atomic arrangement is stronger, causing the Poisson's ratio to be strengthened along the NPR direction. Along the 0° direction, Wave-graphene exhibits almost 0PR behavior, while along the 45° direction it reaches the peak of NPR, ~0.25. This significant difference comes from the fact that the buckling structure activates the reentry mechanism in the direction 45° apart, activating the NPR behavior of Wave-graphene.

Further, we derived the orthogonal response strains along the 0° and 45° directions through strain engineering, as shown in Fig. 4(e). Further, we derived the orthogonal response strains along the 0° and 45° directions through strain engineering, as shown in Fig. 4(e). Along the 0° direction, the response strain in the orthogonal direction has a weak correlation with the forced strain, indicating that the lattice constant hardly changes with strain, *i.e.*, 0PR. Along the 45° direction, the response strain and forced strain show a strong positive linear correlation, indicating that the corresponding strain increases with the increase of the lattice constant, that is, NPR. Thus, Poisson's ratio can be derived by taking the negative value of the slope of the response strain ($s_R$) versus the forced strain ($s_F$):

$$\nu = -s_R/s_F \,, \qquad (12)$$

In the 0° direction, Poisson's ratio almost exhibits 0PR behavior and 25° exhibits NPR behavior exceeding -0.2, which is almost consistent with the results of Equ. (12), demonstrating the accuracy and self-consistency of calculations on different methods.

From a microscopic level, the expansion behavior of the lattice can be naturally related to the vibrational properties of the lattice. In thermal vibration, the lattice structure under harmonic vibration is periodic vibration around the equilibrium position[55]. These vibrations are reversible and do not permanently change the lattice structure. anharmonic vibration introduces more nonlinear effects into the crystal lattice, making the vibration behavior more complex. As shown in Fig. 4(g), the strain potential energy density along the 0° and 45° directions is fitted:

$$E = A + Bs_F + Cs_F{}^2 + Ds_F{}^3, \qquad (13)$$

Where A, B, C and D are 8.700 (8.700), 0.014 (0.009), 7.718 (9.210) and, -0.835 (-4.084) respectively in the 0° (45°) direction. Significant differences occur in the higher order coefficients, that is, the values of D, where the D value along the 45° direction is almost five times higher than that along the 0° direction. Higher coefficient values in the anharmonic term mean strong anharmonic interactions, which indicates that the carbon atoms in Wave-graphene prefer to change along 45° in atomic vibrations. We further quantify this behavior using the Grüneisen parameter (γ), which measures the sensitivity of the phonon vibration frequency in the material to changes in volume. As shown in Fig. 4(h), the γ value changes along the Γ-M and Γ-K directions exhibit obvious differences, indicating its anisotropic lattice vibration properties. When stretched, the lattice response along the 45% direction is more obvious, showing a stronger expansion response while the 0° direction is relatively weak.

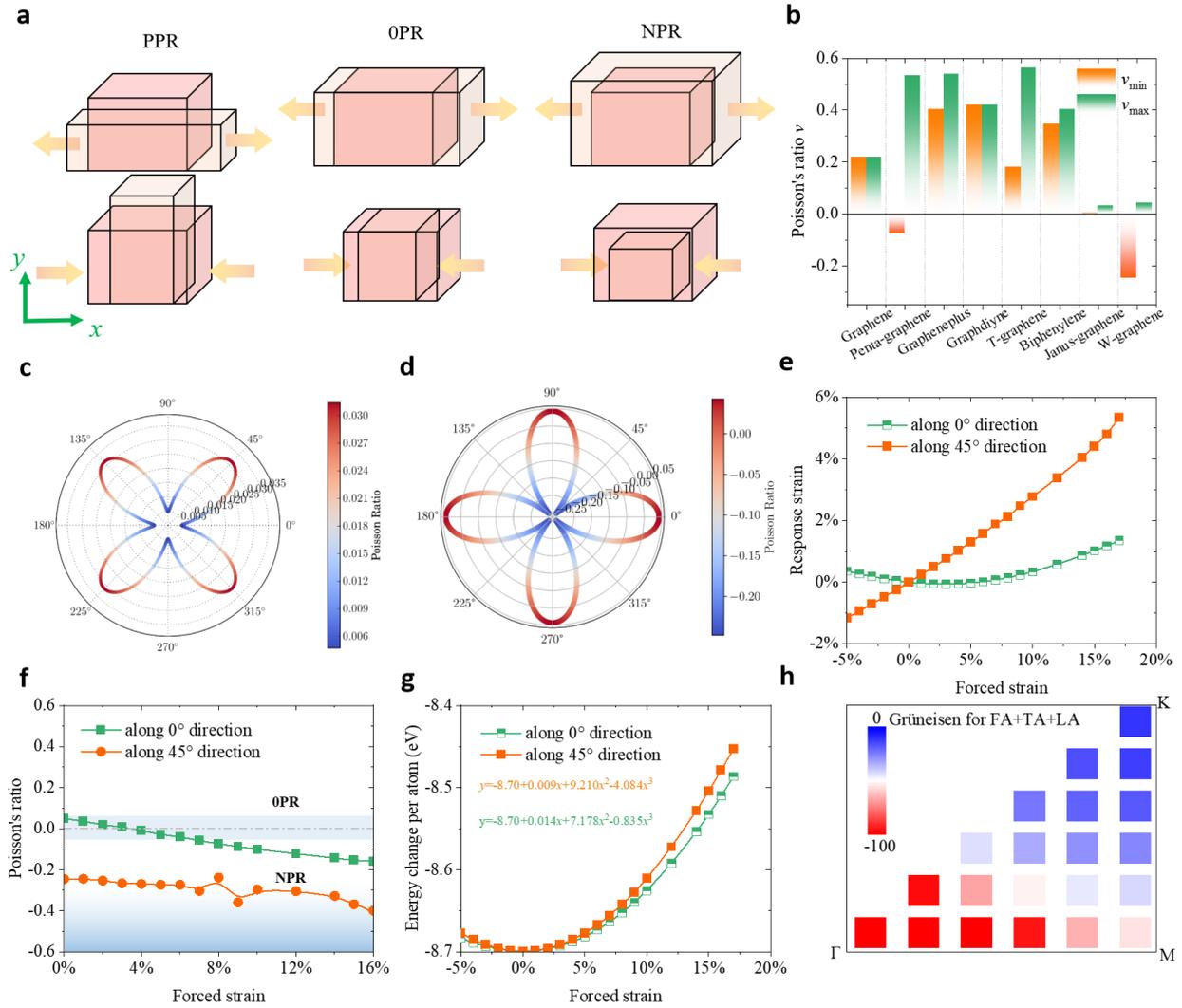

Figure 4. **Anisotropic Poisson's ratio behavior**. (a) Schematic diagram of positive (PPR), zero Poisson's ratio (0PR), and negative Poisson's ratio (NPR) behavior. (b) The maximum and minimum Poisson's ratios of eight carbon allotropes calculated based on Equ. (10). (c) Poisson's ratio in different directions for Janus-graphene. (d) Poisson's ratio in different directions for Wave-graphene. (e) The dependence of response strain on forced strain. (f) The dependence of Poisson's ratio on forced strain. (g) The dependence of system energy variation on forced strain. (h) The dependence of Grüneisen Parameters on forced strain.

**Conclusion**

In summary, we propose a pure $sp^2$ hybrid high buckling carbon allotrope (Wave-graphene) in this work. As a derivative phase of Janus-graphene, it has lower energy. We conducted an in-depth investigation of its geometric, electrical, optical, and mechanical characteristics, and made a comparative analysis with seven distinct typical 2D carbon allotropes. Wave-graphene has a pure $sp^2$ hybrid and high buckling geometric structure, consisting of 4-6-8 carbon rings. In terms of electricity, we have revealed its quasi-direct semiconductor properties with broadband gaps and excellent transparent optical properties. Importantly, Wave-graphene exhibits the anisotropic full stretching properties, *i.e.*, full-auxetic behavior across NPR and 0PR. This anisotropic behavior is due to the activation of the re-entry mechanism caused by high buckling. A deep understanding of the anisotropic nonlinear interactions that lead to the anisotropic response of lattice vibrations can be quantified through potential wells and Grüneisen parameters. The unique full-auxetic behavior, combined with electrical and optical properties, provides valuable insights into the physical properties of Wave-graphene and its potential applications in various fields such as electronics, optoelectronics, and mechanics, and enriches the 2D carbon family.

**Method**

Accurate energy and force calculations are based on density functional theory (DFT) by the Perdew–Burke–Ernzerhof (PBE) [56] functional with the *Vienna ab initio simulation package* (VASP) [55] code. The kinetic energy cutoff of 1000 eV with a 21×21×1 Monkhorst-Pack [58] *q*-mesh was used for structure optimization until the energy and the Hellmann-Feynman force accuracy are $10^{-6}$ eV and $10^{-4}$ eV/Å. The phonon dispersion is calculated based on the finite displacement difference method via the PHONOPY code [59] with the 3×3×1 supercell. *Ab initio molecular dynamics* (AIMD) are calculated based on the canonical ensemble based on VASP code.

**ACKNOWLEDGEMENTS**

This work is supported by the National Natural Science Foundation of China (Grant No. 52006057), the Natural Science Foundation of Chongqing, China (No. CSTB2022NSCQ-MSX0332), the Fundamental Research Funds for the Central Universities (Grant Nos. 531119200237), and the State Key Laboratory of Advanced Design and Manufacturing for Vehicle Body at Hunan University (Grant No. 52175013). H.W. is supported by the National Natural Science Foundation of China (Grant No. 51906097). Z.Q. is supported by the National Natural Science Foundation of China (Grant No.12274374, 11904324) and the China Postdoctoral Science Foundation (2018M642776). The numerical calculations in this paper have been done on the supercomputing system of the E.T. Cluster and the National Supercomputing Center in Changsha.

## AUTHOR CONTRIBUTIONS

*G.Q.* supervised the project. *L.Y.* performed all the calculations, analysis and writing. All the authors contributed to interpreting the results. The manuscript was written by *L.Y.* with contributions from all the authors.

## COMPETING INTERESTS

The Authors declare no Competing Financial or Non-Financial Interests

## DATA AVAILABILITY

The data that support the findings of this study are available from the corresponding author on reasonable request.

## REFERENCES

[1] A. Hashimoto, K. Suenaga, A. Gloter, K. Urita, and S. Iijima, *Direct Evidence for Atomic Defects in Graphene Layers*, Nature **430**, 870 (2004).

[2] Center for Applied Physics and Technology, College of Engineering, Peking University, Beijing 100871, China et al., *PENTA-GRAPHENE: A NEW CARBON ALLOTROPE*, RENSIT **7**, 191 (2015).

[3] M. A. Nazir, A. Hassan, Y. Shen, and Q. Wang, *Research Progress on Penta-Graphene and Its Related Materials: Properties and Applications*, Nano Today **44**, 101501 (2022).

[4] Y. Shen and Q. Wang, *Pentagon-Based 2D Materials: Classification, Properties and Applications*, Physics Reports **964**, 1 (2022).

[5] Y. Hu et al., *Synthesis of γ-Graphyne Using Dynamic Covalent Chemistry*, Nat. Synth **1**, 6 (2022).

[6] C. Huang, Y. Li, N. Wang, Y. Xue, Z. Zuo, H. Liu, and Y. Li, *Progress in Research into 2D Graphdiyne-Based Materials*, Chem. Rev. **118**, 7744 (2018).

[7] X. Gao, H. Liu, D. Wang, and J. Zhang, *Graphdiyne: Synthesis, Properties, and Applications*, Chemical Society Reviews **48**, 908 (2019).

[8] L. Yu, Z. Qin, H. Wang, X. Zheng, and G. Qin, *Half-Negative Poisson's Ratio in Graphene+ with Intrinsic Dirac Nodal Loop*, Cell Reports Physical Science **3**, 100790 (2022).

[9] L. Yu, A. Chen, X. Wang, H. Wang, Z. Qin, and G. Qin, *Softened $sp2-sp3$ Bonding Network Leads to Strong Anharmonicity and Weak Hydrodynamics in Graphene+*, Phys. Rev. B **106**, 125410 (2022).

[10] Y. Liu, G. Wang, Q. Huang, L. Guo, and X. Chen, *Structural and Electronic Properties of $T$ Graphene: A Two-Dimensional Carbon Allotrope with Tetrarings*, Phys. Rev. Lett. **108**, 225505 (2012).

[11] H. P. Veeravenkata and A. Jain, *Density Functional Theory Driven Phononic Thermal Conductivity Prediction of Biphenylene: A Comparison with Graphene*, Carbon **183**, 893 (2021).

[12] D. Ferguson, D. J. Searles, and M. Hankel, *Biphenylene and Phagraphene as Lithium Ion Battery Anode Materials*, ACS Appl. Mater. Interfaces **9**, 20577 (2017).

[13] Q. Fan et al., *Biphenylene Network: A Nonbenzenoid Carbon Allotrope*, Science **372**, 852 (2021).

[14] L. Hou, X. Cui, B. Guan, S. Wang, R. Li, Y. Liu, D. Zhu, and J. Zheng, *Synthesis of a Monolayer Fullerene Network*, Nature **606**, 7914 (2022).

[15] F. Pan et al., *Long-Range Ordered Porous Carbons Produced from C60*, Nature **614**, 7946 (2023).

[16] E. Meirzadeh et al., *A Few-Layer Covalent Network of Fullerenes*, Nature **613**, 7942 (2023).

[17] C.-T. Toh et al., *Synthesis and Properties of Free-Standing Monolayer Amorphous Carbon*, Nature **577**, 7789 (2020).

[18] Q. Hou, J. Wang, and Y. Xiong, Research progress on the improvement of hydrogen storage performance of MgH2 by carbon materials, Rare Met, 47,2023, 47pp. 1614-1623.

[19] Yang, Y., Duan, XC., Guo, SH. et al. Crystalline-amorphous M@MNx (M = Co, Fe, Ni) encapsulated in nitrogen-doped carbon for enhanced efficient and durable hydrogen evolution reaction. Rare Met. (2023).

[20] Wang, LP., Li, K., Ding, HL. et al. Honeycomb-like MoCo alloy on 3D nitrogen-doped porous graphene for efficient hydrogen evolution reaction. Rare Met. (2023).

[21] L. D. Landau, E. M. Lifshitz, R. Atkin, and N. Fox, *The Theory of Elasticity*, in *Physics of Continuous Media* (CRC Press, 2020), pp. 167–178.

[22] A. W. Lipsett and A. I. Beltzer, *Reexamination of Dynamic Problems of Elasticity for Negative Poisson's Ratio*, The Journal of the Acoustical Society of America **84**, 2179 (1988).

[23] J. B. Choi and R. S. Lakes, *Fracture Toughness of Re-Entrant Foam Materials with a Negative Poisson's Ratio: Experiment and Analysis*, Int J Fract **80**, 73 (1996).

[24] R. S. Lakes and K. Elms, *Indentability of Conventional and Negative Poisson's Ratio Foams*, Journal of Composite Materials **27**, 1193 (1993).

[25] S. Blackburn and D. I. Wilson, *Shaping Ceramics by Plastic Processing*, Journal of the European Ceramic Society **28**, 1341 (2008).

[26] Q. Liu, *Literature Review: Materials with Negative Poisson's Ratios and Potential Applications to Aerospace and Defence*, (2006).

[27] J. B. CHOI and R. S. LAKES, *Design of a Fastener Based on Negative Poisson's Ratio Foam*, Cell. Polym **10**, 205 (1991).

[28] F. Scarpa, *Auxetic Materials for Bioprostheses [In the Spotlight]*, IEEE Signal Processing Magazine **25**, 128 (2008).

[29] J.-W. Jiang and H. S. Park, *Negative Poisson's Ratio in Single-Layer Black Phosphorus*, Nat Commun **5**, 4727 (2014).

[30] L. Yu, Q. Yan, and A. Ruzsinszky, *Negative Poisson's Ratio in 1T-Type Crystalline Two-Dimensional Transition Metal Dichalcogenides*, Nat Commun **8**, 15224 (2017).

[31] L. Yu, Y. Wang, X. Zheng, H. Wang, Z. Qin, and G. Qin, *Emerging Negative Poisson's Ratio Driven by Strong Intralayer Interaction Response in Rectangular Transition Metal Chalcogenides*, Applied Surface Science **610**, 155478 (2023).

[32] L. Zhang, C. Tang, and A. Du, *Two-Dimensional Vanadium Tetrafluoride with Antiferromagnetic Ferroelasticity and Bidirectional Negative Poisson's Ratio*, Journal of Materials Chemistry C **9**, 95 (2021).

[33] R. Peng, Y. Ma, Z. He, B. Huang, L. Kou, and Y. Dai, *Single-Layer Ag2S: A Two-Dimensional Bidirectional Auxetic Semiconductor*, Nano Lett. **19**, 1227 (2019).

[34] Z. Li, Y. Gu, C. He, and X. Zou, *Fully Auxetic and Multifunctional of Two-Dimensional $\delta$-GeS and $\delta$-GeSe*, Phys. Rev. B **106**, 035426 (2022).

[35] Z. Gao, Q. Wang, W. Wu, Z. Tian, Y. Liu, F. Ma, Y. Jiao, and S. A. Yang, *Monolayer ${\mathrm{RhB}}_{4}$: Half-Auxeticity and Almost Ideal Spin-Orbit Dirac Point Semimetal*, Phys. Rev. B **104**, 245423 (2021).

[36] F. Ma, Y. Jiao, W. Wu, Y. Liu, S. A. Yang, and T. Heine, *Half-Auxeticity and Anisotropic Transport in Pd Decorated Two-Dimensional Boron Sheets*, Nano Lett. **21**, 2356 (2021).

[37] Y. Wu et al., *Three-Dimensionally Bonded Spongy Graphene Material with Super Compressive Elasticity and near-Zero Poisson's Ratio*, Nat Commun **6**, 1 (2015).

[38] H. Yang and L. Ma, *Multi-Stable Mechanical Metamaterials with Shape-Reconfiguration and Zero Poisson's Ratio*, Materials & Design **152**, 181 (2018).

[39] J. N. Grima, L. Oliveri, D. Attard, B. Ellul, R. Gatt, G. Cicala, and G. Recca, *Hexagonal Honeycombs with Zero Poisson's Ratios and Enhanced Stiffness*, Advanced Engineering Materials **12**, 855 (2010).

[40] J. Liu, S. Liu, J.-Y. Yang, and L. Liu, *Electric Auxetic Effect in Piezoelectrics*, Phys. Rev. Lett. **125**, 197601 (2020).

[41] Xu, YX., Fan, HZ. & Zhou, YG. Quantifying spectral thermal transport properties in framework of molecular dynamics simulations: a comprehensive review. Rare Met. 42, 3914–3944 (2023).

[42] Q. K. Tian, P. X. Li, J. H. Wei, Z. Y. Xing, G. Z. Qin, and Z. Z. Qin, *Inverse Janus design of two-dimensional Rashba semiconductors.* Physical Review B **10**, 1103 (2023).

[43] Wang, LP., Li, K., Ding, HL. et al. Honeycomb-like MoCo alloy on 3D nitrogen-doped porous graphene for efficient hydrogen evolution reaction. Rare Met. (2023).

[44] A. D. Becke and K. E. Edgecombe, *A Simple Measure of Electron Localization in Atomic and Molecular Systems*, The Journal of Chemical Physics **92**, 5397 (1990).

[45] P. X. Li, X. Wang, H. Y. Wang, Q. K. Tian, J. Y. Xu, L. F. Yu, G. Z. Qin, and Z. Z. Qin, *Biaxial strain modulated electronic structures of layered two-dimensional MoSiGeN4 Rashba systems*，Phys. Chem. Chem. Phys.,2024, 26, 1891.

[46] Yu, LF., Xu, JY., Shen, C. et al. Realizing ultra-low thermal conductivity by strong synergy of asymmetric geometry and electronic structure in boron nitride and arsenide. Rare Met. 42, 210–221 (2023).

[47] Peng, QQ., Wang, YT., Qi, S. et al. The homogenous growth of Co-based coordination compound on graphene nanosheet for high-performance K-organic battery and its reaction mechanism. Rare Met. (2024).

[48] Wei, DH., Zhou, E., Xu, JY. et al. Insight into vertical piezoelectric characteristics regulated thermal transport in van der Waals two-dimensional materials. Rare Met. (2023).

[49] H. Wang, G. Qin, J. Yang, Z. Qin, Y. Yao, Q. Wang, and M. Hu, *First-Principles Study of Electronic, Optical and Thermal Transport Properties of Group III–VI Monolayer MX (M = Ga, In; X = S, Se)*, Journal of Applied Physics **125**, 245104 (2019).

[50] M. Gajdoš, K. Hummer, G. Kresse, J. Furthmüller, and F. Bechstedt, *Linear Optical Properties in the Projector-Augmented Wave Methodology*, Phys. Rev. B **73**, 045112 (2006).

[51] J. Y. Yang and L. H. Liu, *Effects of Interlayer Screening and Temperature on Dielectric Functions of Graphene by First-Principles*, Journal of Applied Physics **120**, 034305 (2016).

[52] H. Bao and X. Ruan, *Ab Initio Calculations of Thermal Radiative Properties: The Semiconductor GaAs*, International Journal of Heat and Mass Transfer **53**, 1308 (2010).

[53] M. Maździarz, *Comment on `The Computational 2D Materials Database: High-Throughput Modeling and Discovery of Atomically Thin Crystals'*, 2D Mater. **6**, 048001 (2019).

[54] V. Wang, Y.-Y. Liang, Y. Kawazoe, and W.-T. Geng, *High-Throughput Computational Screening of Two-Dimensional Semiconductors*, arXiv:1806.04285 [Cond-Mat] (2020).

[55] Liu, ZK., Shen, Y., Li, HL. et al. Observation of ballistic-diffusive thermal transport in GaN transistors using thermoreflectance thermal imaging. Rare Met. 43, 389–394 (2024).

[56] J. P. Perdew, K. Burke, and M. Ernzerhof, *Generalized Gradient Approximation Made Simple*, Phys. Rev. Lett. **77**, 3865 (1996).

[57] G. Kresse and J. Hafner, Ab Initio *Molecular-Dynamics Simulation of the Liquid-Metal–Amorphous-Semiconductor Transition in Germanium*, Phys. Rev. B **49**, 14251 (1994).

[58] H. J. Monkhorst and J. D. Pack, *Special Points for Brillouin-Zone Integrations*, Phys. Rev. B **13**, 5188 (1976).

[59] L. Chaput, A. Togo, I. Tanaka, and G. Hug, *Phonon-Phonon Interactions in Transition Metals*, Phys. Rev. B **84**, 094302 (2011).